\documentclass[twocolumn,amssymb, pra]{revtex4}
\usepackage{graphicx}

\begin{document}
\title{Initial state dependence of a quantum-resonance ratchet}

\author{Jiating Ni,$^{1}$ Wa Kun Lam,$^{1}$ Siamak Dadras,$^{1}$ Mario F. Borunda,$^{1}$ Sandro Wimberger,$^{2,3,4}$ and Gil S. Summy$^{1}$}

\affiliation{$^1$Department of Physics, Oklahoma State University, Stillwater, Oklahoma 74078-3072, USA}
\affiliation{$^2$DiFeST, Universit$\grave{a}$ degli Studi di Parma, Parco delle Scienze n. 7/A, 43124 Parma, Italy}
\affiliation{$^3$INFN, Sezione di Milano Bicocca, Gruppo Collegato di Parma, Parma, Italy}
\affiliation{$^4$ITP, Heidelberg University, Philosophenweg 12, 69120 Heidelberg, Germany}

\date{\today}%

\begin{abstract}
We demonstrate quantum resonance ratchets created with Bose-Einstein condensates exposed to pulses of an off-resonant standing light wave. We show how some of the basic properties of the ratchets are controllable through the creation of different initial states of the system. In particular, our results prove that through an appropriate choice of initial state it is possible to reduce the extent to which the ratchet state changes with respect to time. We develop a simple theory to explain our results and indicate how ratchets might be used as part of a matter wave interferometer or quantum-random walk experiment.



\end{abstract}
\maketitle

\section{Introduction}

Understanding the nature of quantum transport is an important problem for quantum chaos and for many condensed matter systems \cite{39}. The delta-kicked rotor is an ideal system for studying quantum transport and can be realized by exposing a sample of cold atoms to short pulses of an optical standing wave. This is the so-called atom optics quantum kicked rotor (AOQKR) \cite{1}. The implementation of the AOQKR in experiments was a breakthrough in the study of quantum chaos, and the system has now been widely used in elucidating phenomena like dynamical localization \cite{1}, quantum resonances (QRs) \cite{1,2,3}, quantum accelerator modes (QAMs) \cite{4,5,6,7,8,27}, fidelity \cite{9,10,11,12}, and quantum ratchets \cite{13,14,15,16,17,21,22,23,29,42}. In contrast to Brownian ratchets \cite{18}, quantum ratchets do not experience dissipative processes or a net biasing force \cite{44}. Hence these systems have been proposed as the basis for a Hamiltonian quantum ratchet \cite{16}. In previous experimental observations of quantum ratchets \cite{14,18, 24}, it was found that only a relatively small component of the initial wave function contributed to the ratchet current. It is therefore natural to ask if it is possible to improve the mode matching between the initial state of a system and the ratchet state so as to increase the participation in the ratchet. This could be important for applications in atom optics such as interferometry. Our motivation for enhancing the efficiency of the quantum ratchet is the possible use of such a system as the walk component in experiments on quantum random walks \cite{25,31,43}.

The structure of this paper is as follows; we begin by providing a theoretical analysis of the intrinsic mechanism of a quantum ratchet. This includes the development of a simple classical picture that can explain many of the features of the ratchet. We then present the experimental realization of the momentum currents using initial states that have been prepared using Bragg diffraction with light pulses. We demonstrate how different phases on the initial state wave function can affect the momentum current and examine how the momentum current depends on the number of momentum states in the initial wave function. In particular, we present results showing a connection between the number of momentum state components and the ``dispersion'' of the ratchet.

\section{Theory}
We start by summarizing the basic dynamics of the AOQKR system. The Hamiltonian in dimensionless units is given by \cite{38}:
\begin{equation}
\label{ }
\hat{H}=\frac{\hat{p}^{2}}{2} + \phi_{d}V(\hat{x})\sum_{t=1}^{N}\delta(t^{\prime}-t\tau),
\end{equation}
where $\hat{p}$ is the scaled momentum in units of $\hbar G$ (two photon recoils), $\phi_{d}$ is the strength of the kicks, and is given by $\phi_{d}=\Omega^{2}\triangle t/8\delta_{L}$, where $\triangle t$ is pulse length, $\Omega=\vec{\mu}\cdot\vec{E}(\vec{r})/\hbar$ is the Rabi frequency, and $\delta_{L}$ is the detuning of the kicking laser from the atomic transition. $V(\hat{x}) = \cos(Gx+\gamma)$ is the potential created by a standing wave with a grating vector $G=2\pi/\lambda_{G}$. $\lambda_{G}$ is the spatial period of the standing wave, and is determined by the direction of the laser beams and their wavelength. Here, $x$ is position, and $\gamma$ is an offset phase. $t^{\prime}$ is the continuous time variable, and $t$ counts the number of kicks. $\tau=2\pi T / T_{1/2}$ is the scaled pulse period where $T$ is the real pulse period and $T_{1/2}=2\pi M / \hbar G^{2}$ is the half-Talbot time for an atom with mass $M$. The momentum is changed in quanta of $\hbar G$ in this system, and as a result $p$ can be broken down as $p=n+\beta$, where $n$ and $\beta$ are the integer and fractional parts of the momentum respectively. $\beta$, known as the quasi-momentum, is a conserved quantity \cite{36}.

A convenient picture of the ratchet mechanism can be derived by considering the gradient of the standing wave as the driving force on the wave function. From this standpoint ensuring that the wave function is maximized near positions with larger gradients in the potential should produce a net force and hence the possibility of a ratchet. In this picture the sign of the potential gradient near the wave function's maxima will determine the ratchet's direction. One way to create a spatially non-uniform atomic wave function is to use an initial state that is comprised of a superposition of two or more plane waves: $|\psi\rangle=\sum_{n}e^{in\pi/2}|n\rangle$, where $|n\rangle$ refers to the state $|n\hbar G\rangle$. As will be seen shortly, the phase prefactors shift the maxima of the spatial wave function to where the potential gradient is greatest.

Considering the case of a single Bose-Einstein condensate (BEC) for the initial state, the wave function in momentum space can be written as $\psi(p)=\delta(p)$ assuming the BEC has a narrow momentum width. Here $p$ is the continuous momentum variable. In order to study the wave function in the frame of the standing wave, we transform it into position space by using a Fourier transformation. The magnitude of the wave function in position space is $|\phi(x)|=\sqrt{G/2\pi}$. Since this is uniform in $x$, the average force is zero and according to the simple picture described above no ratchet is formed.

When the initial state contains two or more plane waves, the wave function in position space can be written as:
\begin{equation}
\label{ }
\phi(x)=A\sum_{n}e^{-in\pi/2}e^{ip_{n}x/\hbar},
\end{equation}
where $A$ is the normalization factor. By plotting $\left|\phi(x)\right|^{2}$ and the standing wave potential together, it is noticeable that peaks of the wave function arise at positions where the gradient of the standing wave happens to be the greatest (see FIG. 1).  Naturally, the more plane waves composing the initial state, the more localized the wave function should become. In FIG. 1, the bold solid line is the standing wave, the dashed and dotted lines are the wave functions for the initial states with a superposition of seven plane waves: $\sum_{n=-3}^{3}e^{-in\pi/2}|n\rangle$, and two plane waves: $|0\rangle+e^{i\pi/2}|1\rangle$ respectively. The dashed line is much taller and has a reduced full width at half maximum (FWHM). Figure 2 shows that as the number of consecutive plane waves in the initial state increases, the corresponding FWHM of the wave function decreases. Since a peak with larger FWHM experiences a smaller overall gradient and a larger variation in the gradient, we expect that a ``cleaner" ratchet will require more plane waves in the initial superposition. The effective force $F_{\textrm{eff}}=\int_{-\pi}^{\pi}\left|\phi(x)\right|^{2} dV(x)$ of the standing wave (integrate the absolute square of wave function with the standing wave gradient) for different initial states was also calculated. The dashed line in FIG. 2 shows that $F_{\textrm{eff}}$ increases with the number of the plane waves in the initial state.

\begin{figure}
\begin{center}
\includegraphics[width=0.5\textwidth]{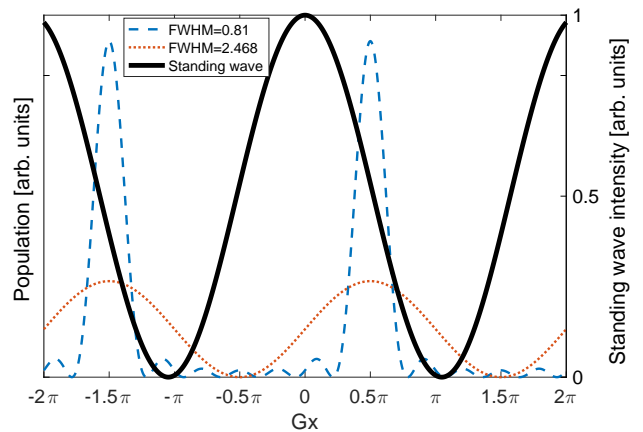}
\caption{The solid line is the standing wave intensity for potential $V(x)$. The dashed line is the wave function for a superposition of seven plane waves: $\sum_{n=-3}^{3}e^{-in\pi/2}|n\rangle$. The dotted line is the wave function for a superposition of two plane waves: $|0\rangle+ e^{i\pi/2}|1\rangle$.}
\label{ }
\end{center}
\end{figure}

\begin{figure}
\begin{center}
\includegraphics[width=0.5\textwidth]{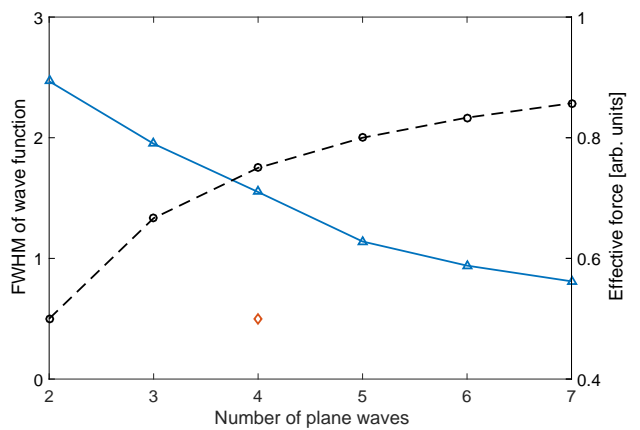}
\caption{The solid line is the theoretical FWHM of the wave function in position space vs number of plane waves, while the dashed line is the theoretical effective force of the standing wave. The diamond is the magnitude of effective force for the initial state $e^{-i\pi}|-2\rangle+e^{-i\pi/2}|-1\rangle+e^{i\pi/2}|1\rangle+e^{i\pi}|2\rangle$. Note that in this case the $n=0$ state is missing.}
\label{ }
\end{center}
\end{figure}

To ensure the peak of the wave function appears at a position that will maximize  $F_{\textrm{eff}}$ (in this case $Gx=\pi/2$), the phase for each plane wave is extremely important. For this reason the phases should be set to $e^{in\pi/2}$, where $n$ is the momentum state index. If the phases differ from these values the peak of the wave function in position space will shift away from the greatest gradient and the effective force will become weaker. We also point out that to maximize the ratchet, the momentum states in the initial superposition should be consecutive. In other words, the momentum difference between neighboring momentum states should be $\hbar G$. Figure 3 shows the wave function in position space for initial states with a superposition of nonconsecutive momentum states. Note that besides the wave function peak at the greatest gradient point, there is also considerable wave function amplitude at positions where the gradient is zero or of opposite sign. This can provide an explanation for why the ratchet is weak or even absent in experiments with these initial states. The dashed line represents the initial state: $e^{-i\pi}|-2\rangle+e^{-i\pi/2}|-1\rangle+e^{i\pi/2}|1\rangle+e^{i\pi}|2\rangle$ and shows a relatively large peak at the greatest gradient point. Thus the ratchet effect should still exist, although is weaker than the ratchet from an initial state with four consecutive momentum states. This can be seen in FIG. 2 where the diamond indicates $F_{\textrm{eff}}$ for this state. In FIG. 3 the dotted line represents the initial state: $e^{-i\pi}|-2\rangle+e^{i\pi}|2\rangle$. This has $F_{\textrm{eff}}=0$. Finally, the dash-dot line is for the initial state: $e^{-i\pi/2}|-1\rangle+e^{i\pi/2}|1\rangle$. Note that two wave function peaks appear at points where the positive and negative gradient is greatest on the standing wave. This suggests that we should expect two simultaneously ratchets with opposite directions.
\begin{figure}
\begin{center}
\includegraphics[width=0.5\textwidth]{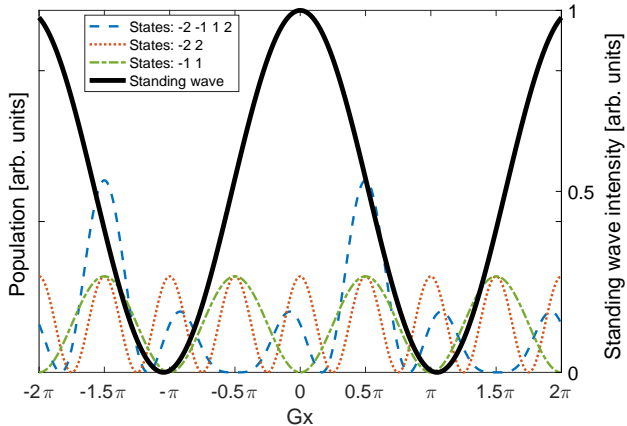}
\caption{The solid line is the standing wave intensity for potential $V(x)$. The dashed line is the wave function for a superposition of nonconsecutive plane waves: $e^{-i\pi}|-2\rangle+e^{-i\pi/2}|-1\rangle+e^{i\pi/2}|1\rangle+e^{i\pi}|2\rangle$. The dotted line is the wave function for a superposition of nonconsecutive plane waves: $e^{-i\pi}|-2\rangle+e^{i\pi}|2\rangle$. The dash-dot line is the wave function for a superposition of nonconsecutive plane waves: $e^{-i\pi/2}|-1\rangle+e^{i\pi/2}|1\rangle$.}
\label{ }
\end{center}
\end{figure}

\section{Experimental setup}
Our experimental apparatus is similar to the one described in Ref. [24]. A BEC with about 70,000 $^{87}$Rb atoms was created in the $5S_{1/2}$, $F=1$ hyperfine ground states in a focused CO$_{2}$ laser beam using an all-optical trapping technique \cite{37}. The ratio of the mean field energy to the recoil energy is approximatively $10^{-5}$, which puts us in the region where the nonlinearity has a negligible effect on the dynamics. Immediately after the BEC was released from the trap, we applied a series of horizontal optical standing wave pulses. The standing wave was formed by two laser beams with a wavelength of $\lambda=780$ nm, which was 6.8 GHz red detuned from the $5S_{1/2}$, $F=1$ transition $\longrightarrow$ $5P_{3/2}$, $F'=3$ transition. Each laser beam was aligned $53^{\circ}$ from the vertical to give a standing wave spacial period of $\lambda_{G}=\lambda/(2\sin53^{\circ})$. This led to a half-Talbot time of  $T_{1/2}\approx 51.5$ $\mu$s, with the primary QRs falling at integer multiples of this time \cite{2}. To create the necessary standing wave pulses, we controlled the phase, intensity, pulse length, and the relative frequency between the two laser beams. This was achieved by passing each of the standing wave's constituent laser beams through an acousto-optic modulator (AOM) driven by an arbitrary waveform generator. The nodes of the standing wave moved with a velocity given by $v=2\pi\triangle f/G$ with $\triangle f$ being the frequency difference between the two beams. The momentum of the BEC, $p$, relative to the standing wave is proportional to $v$ and therefore $p$ can be controlled through $\triangle f$. The kicking pulse length was fixed at 600 $n$s to ensure that the experiments were performed in the Ramam-Nath regime \cite{40,41}. In other words, the evolution of the wave function due to its kinetic energy is ignored during the pulse.

To prepare an initial state comprised of a superposition of several plane waves, a sequence of longer standing wave pulses were applied in the Bragg configuration \cite{28}. Such pulses connect two momentum states with an interaction matrix given by
\begin{equation}
\label{ }
U=\left(\begin{array}{cc}
  \cos(\frac{\Omega_{B}\tau_{B}}{2})    &  -i\sin(\frac{\Omega_{B}\tau_{B}}{2})\exp(i\gamma_{B}) \\[3mm]
  -i\sin(\frac{\Omega_{B}\tau_{B}}{2})\exp(-i\gamma_{B})    &   \cos(\frac{\Omega_{B}\tau_{B}}{2})
\end{array}\right),
\end{equation}
where $\Omega_{B}$ is the effective Rabi frequency, $\tau_{B}$ is the Bragg pulse length, and $\gamma_{B}$ is the offset phase of the standing wave used for a Bragg pulse. To understand how Bragg pulses were utilized in our experiment, consider the preparation of a superposition of five momentum states. Figure 4 shows the procedure for creating this superposition. In the first setup a Bragg pulse with $p=0.5$ diffracts atoms from $|0\rangle$ to $|1\rangle$. The second Bragg pulse with $p=-0.5$ diffracts atoms from $|0\rangle$ to $|-1\rangle$ without affecting the atoms in state $|1\rangle$. Similarly, the third and fourth Bragg pulses, with $p=1.5$ and $p=-1.5$, diffract atoms from $|1\rangle$ to $|2\rangle$ and from $|-1\rangle$ to $|-2\rangle$ respectively. All of the Bragg pulses were applied consecutively without dwell time between the pulses.  There is considerable freedom in the choice of pulse length, however in our case this was set at 103 $\mu$s or $1\times$Talbot time. This ensured that the free evolution of each of the prepared states was always $\hat{1}$ during the application of subsequent Bragg pulses. The intensity for the different pulses was carefully adjusted to make the population in each state equal. We implemented preparation schemes like this for superpositions comprising up to 7 states. The relative phases between the Bragg prepared states are critically important to the dynamics of the ratchet. We performed the experiments so that all of the phases of the states in the superposition were identical. This was achieved by setting $\gamma_{B}=-\pi/2$ or $\gamma_{B}=\pi/2$ depending on whether we were coupling states with $\Delta n=-1$ or $\Delta n =1$ respectively.


\begin{figure}
\begin{center}
\includegraphics[width=0.5\textwidth]{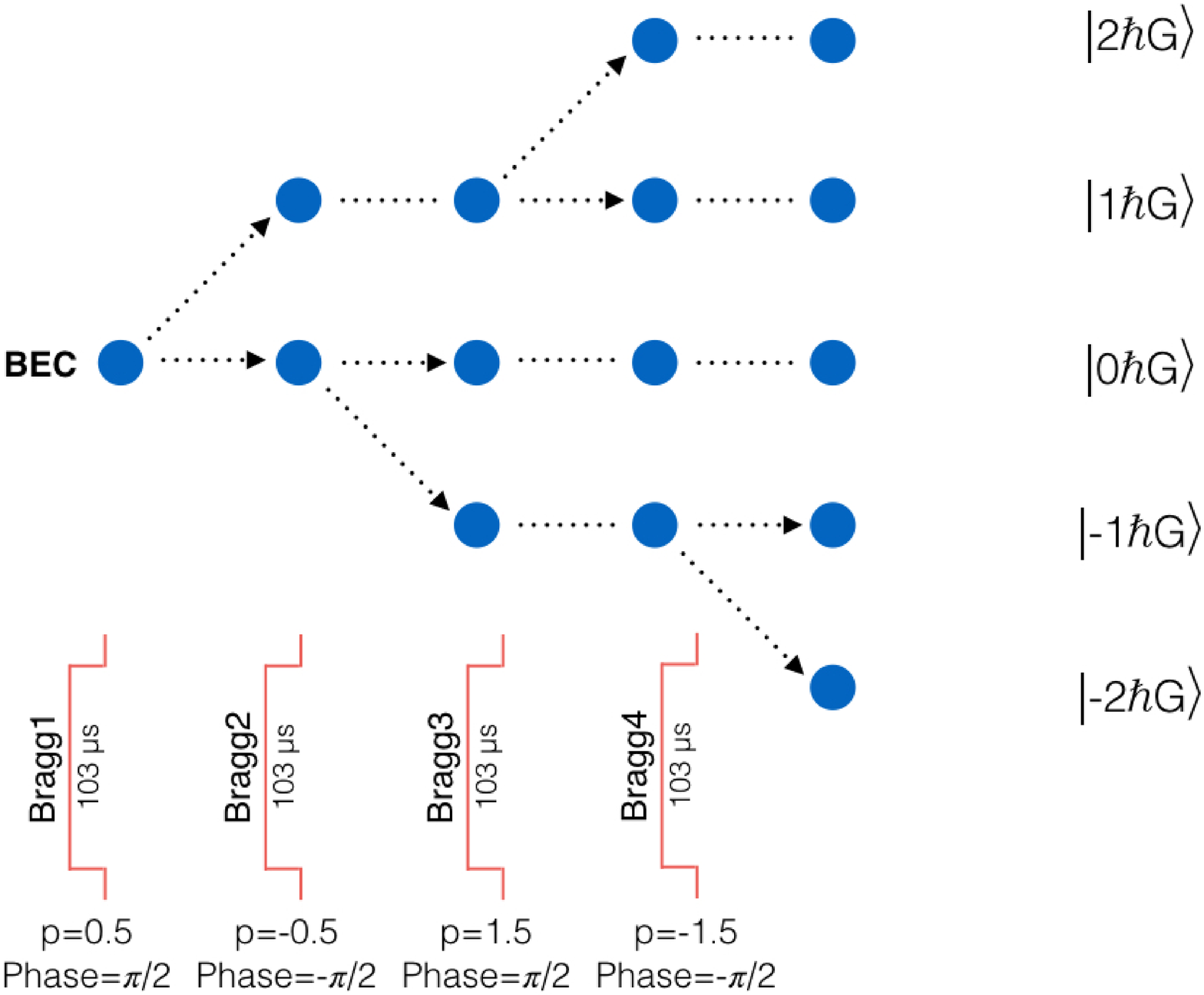}
\caption{Experimental schematic showing preparation of the five component initial state: $\sum_{n=-2}^{2}|n\rangle$. }
\label{ }
\end{center}
\end{figure}

Following the Bragg preparation, the delta-like kick rotor pulses were immediately applied. These diffracted the atoms into a wide range of momentum states. For the kicking pulses, the pulse strength $\phi_{d}$ $\sim$ 1.4, and the phase $\gamma$ in the potential was $\pi/2$. This is mathematically equivalent to individually setting the phases of the initial momentum states to be $e^{in\pi/2}$ which is required to maximize the ratchet. For all but a few experiments, the time between kicking pulses was 51.5 $\mu$s (half-Talbot time) with initial momentum $\beta=0.5$ to maintain QR. In order to measure the momentum distribution after the pulse sequence, atoms were absorption imaged following a free flight time of 9 ms. Several examples of time-of-flight images of the BEC vs $t$ are shown in FIG. 5. Of particular note is the experiment with the initial state $|-1\rangle + |1\rangle$ (FIG. 5(f)). This data highly supports the idea from FIG. 3 that the two identical peaks appearing at the points of greatest positive and negative gradient should produce two ratchets with opposite directions. This effect might be applied in atomic interferometry as a beamsplitter \cite{34}.

\begin{figure}
\begin{center}
\includegraphics[width=0.5\textwidth]{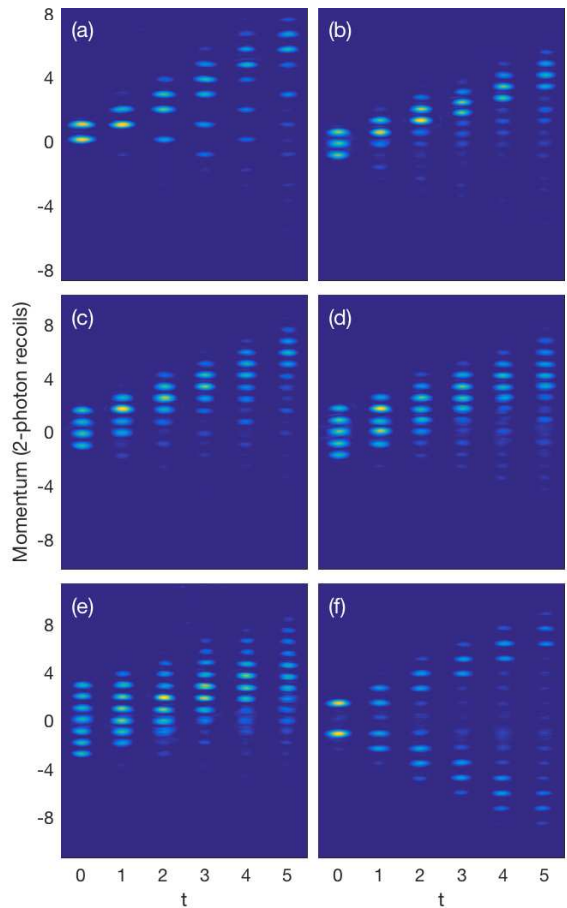} 
\caption{Experimental time-of-flight images as a function of $t$. The relative phases of the momentum components in the initial states were zero. The initial states for panels (a)-(f) were respectively: $|0\rangle+|1\rangle$, $\sum_{n=-1}^{1}|n\rangle$, $\sum_{n=-1}^{2}|n\rangle$, $\sum_{n=-2}^{2}|n\rangle$, $\sum_{n=-3}^{3}|n\rangle$, $|-1\rangle+|1\rangle$. In each case the $t=0$ images show the initial state. Panels (a)-(e) show the single ratchets with positive directed currents. In (f) a symmetric double ratchet is seen. }
\label{ }
\end{center}
\end{figure}

As can be seen from the previous discussion, the phases of the components of the superposition are extremely important for the ratchet dynamics. Hence it is critical that any unwanted phases be controlled and if possible eliminated. To this end we implemented a Mach-Zehnder Bragg interferometer \cite{28} with which we could measure such phases. In particular, we were able to reduce extraneous phase shifts due to gravity (non-horizontal standing wave) and ac-Stark shifts (due to small amounts of stray light) to levels where they were insignificant for the ratchet experiment.

\section{Data analysis and discussion}
A large part of the motivation for the work presented here is to understand the behavior of ratchets with different initial states. To this end we have performed a comprehensive experimental study of the ``dispersion'' of a ratchet as a function of several variables. We define the normalized dispersion of a ratchet as:
\begin{equation}
\label{ }
D(t)=\frac{\langle p_{t}^{2}\rangle-\langle p_{t}\rangle^{2}}{\langle p_{0}^{2}\rangle-\langle p_{0}\rangle^{2}},
\end{equation}
where $\langle p_{t}\rangle$ is the mean momentum, $\langle p_{t}^{2}\rangle$ is the mean of momentum square, and $t$ is the kick number ($t=0$ corresponds to the initial distribution). The dispersion is an objective way of describing the ``quality" of the ratchet current: the closer $D(t)$ is to $1$ the closer the ratchet state resembles its initial form. Reducing the amount by which the dispersion increases can be important for applications, such as realizing quantum random walks \cite{31}. We now examine the sensitivity of the ratchet current dispersion to the initial state. Figure 6 shows the dependence of the dispersion on kick number $t$ for $\beta=0.5$ and $T=T_{1/2}$. The experimental results are in line with what we would expect by looking at the wave function plots in FIG. 1 and show that a larger number of momentum states in the initial state produces a smaller dispersion. In other words, to improve the quality of the ratchet we need to generate an initial state with a large number of plane waves in its superposition.
\begin{figure}
\begin{center}
\includegraphics[width=0.5\textwidth]{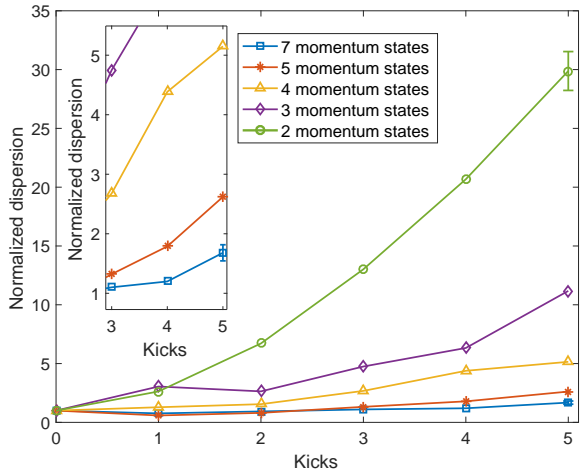}
\caption{Experimental data showing normalized dispersion vs kick number $t$ with $\beta=0.5$ and $T=T_{1/2}$ for initial states $\sum_{n=-3}^{3}|n\rangle$ (squares), $\sum_{n=-2}^{2}|n\rangle$ (asterisks), $\sum_{n=-1}^{2}|n\rangle$ (triangles), $\sum_{n=-1}^{1}|n\rangle$ (diamonds), and $\sum_{n=0}^{1}|n\rangle$ (circles). The inset gives a closer view of the dispersion. Representative error bars are given for: 7 momentum states, kicks=5; 2 momentum states, kicks=5. }
\label{ }
\end{center}
\end{figure}

As discussed before, the ratchet is sensitive to the phases in the initial state. To illustrate this point, experiments were carried out with an initial state $\sum_{n=-2}^{2}|n\rangle$ in which an extra phase $\pi/2$ was added to momentum state $|1\rangle$. Figure 7 shows that the dispersion grows more quickly when the phase of momentum state $|1\rangle$ differs from the optimal value.
\begin{figure}
\begin{center}
\includegraphics[width=0.5\textwidth]{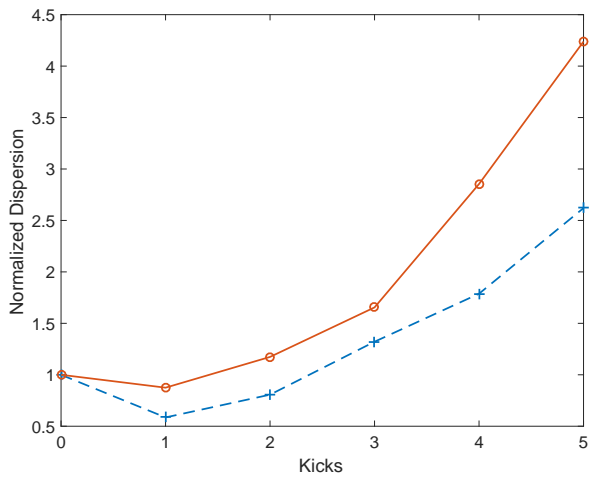}
\caption{Experimental data showing normalized dispersion vs kick number $t$ with $\beta=0.5$ and $T=T_{1/2}$ for initial state $\sum_{n=-2}^{2}|n\rangle$. The dashed line is for an initial state with optimal phase (net phase on each momentum state is zero). The solid line is for an initial state where the phase of momentum state $|1\rangle$ is $\pi/2$. }
\label{ }
\end{center}
\end{figure}

We also investigated the ratchet current using kicks separated by different QR times. In principle, kicking pulses separated by $T=0$ with $\beta=0$ or $T=T_{T}$ (Talbot time) with $\beta=0$ should give the same ratchet current as $T=T_{1/2}$ with $\beta=0.5$. The experimental results show that with the same initial state, dispersion is smallest with $T=0$ and largest with $T=T_{T}$ (see FIG. 8). The reason is probably that phase noise from the environment and de-phasing of the initial state ($\Delta\beta$, the width of the initial state, is small but not $0$ and results in phase-evolution during $T$) are minimized by setting the $T=0$ \cite{30}. This is supported by the observation that the ratchet is sensitive to the phase of the initial state. 

\begin{figure}
\begin{center}
\includegraphics[width=0.5\textwidth]{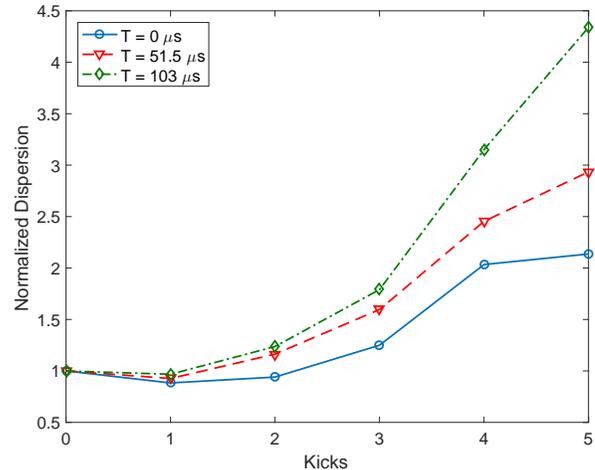}
\caption{Experimental data showing normalized dispersion vs kick number $t$ for initial state $\sum_{n=-2}^{2}|n\rangle$ with the standing wave phase $\gamma=\pi/2$. The solid line is for $\beta=0$ and $T=0$. The dashed line is for $\beta=0.5$ and $T=51.5$ $\mu$s (half-Talbot time). The dash-dot line is for $\beta=0$ and $T=103$ $\mu$s (Talbot time). }
\label{ }
\end{center}
\end{figure}


\section{Conclusions}
Experiments were carried out to create an on-resonant atomic  ratchet by exposing an initial atomic state which was a superposition of two or more momentum states to a series of standing wave pulses. A picture that we used to understand many of the features of the ratchet considers the effective force produced by the standing wave pulses. When more than one plane wave is present in the initial state this force can be non-zero. We defined and measured the dispersion of the momentum of the ratchet current as a function of kick number $t$. It was verified that the ratchet dispersion grew more slowly when the initial state contained a large number of consecutive momentum states. We studied the effect from the phase of the initial state to the ratchet current. A small error in the phase causes a change in the dispersion which is not negligible. We also performed experiments with different times between the kicking pulses and concluded that shorter times between were better so as to minimize perturbations from the environment that cause dephasing of ratchet states. We hope that this work will provide the basis for further studies of ratchets, particularly as they pertain to their application for experiments with quantum random walks.

\section{Acknowledgments}
We thank J. Clark and A. Sundararaj for helpful comments and discussions.

{}


\begin{thebibliography}{}

\bibitem{39}
  C. W. J. Beenakker, Rev. Mod. Phys. \textbf{69}, 731 (1997).
  
  \bibitem{1}
  F. L. Moore, J. C. Robinson, C. Bharucha, P. E. Williams, and M. G. Raizen, Phys. Rev. Lett. \textbf{73}, 2974 (1994).

  \bibitem{2}
  C. Ryu, M. F. Andersen, A. Vaziri, M. B. d'Arcy, J. M. Grossman, K. Helmerson, and W. D. Phillips, Phys. Rev. Lett. \textbf{96}, 160403 (2006).

  \bibitem{3}
  F. M. Izrailev, Phys. Rep. \textbf{196}, 299 (1990).

  \bibitem{4}
  S. Fishman, I. Guarneri, and L. Rebuzzini, Phys. Rev. Lett. \textbf{89}, 084101 (2002).

  \bibitem{5}
  S. Fishman, I. Guarneri, and L. Rebuzzini, J. Stat. Phys. \textbf{110}, 911 (2003).

  \bibitem{6}
  G. Behinaein, V. Ramareddy, P. Ahmadi, and G. S. Summy, Phys. Rev. Lett. \textbf{97}, 244101 (2006).

  \bibitem{7}
  V. Ramareddy, G. Behinaein, I. Talukdar, P. Ahmadi, and G. S. Summy, EPL (Europhysics Letters) \textbf{89}, 33001 (2010).

  \bibitem{8}
  M. K. Oberthaler, R. M. Godun, M. B. d'Arcy, G. S. Summy, and K. Burnett, Phys. Rev. Lett. \textbf{83}, 4447 (1999).

  \bibitem{27}
  R. K. Shrestha, J. Ni, W. K. Lam, G. S. Summy, and S. Wimberger, Phys. Rev. E \textbf{88}, 034901 (2013).


  \bibitem{9}
  P. McDowall, A. Hilliard, M. McGovern, T. Gr\"{u}nzweig, and M. F. Andersen, New J. Phys. \textbf{11}, 123021 (2009).

  \bibitem{10}
  A. Ullah and M. D. Hoogerland, Phys. Rev. E \textbf{83}, 046218 (2011).

  \bibitem{11}
  R. K. Shrestha, S. Wimberger, J. Ni, W. K. Lam, and G. S. Summy, Phys. Rev. E \textbf{87}, 020902(R) (2013).

  \bibitem{12}
  I. Talukdar, R. Shrestha, and G. S. Summy, Phys. Rev. Lett \textbf{105}, 054103 (2010).

  \bibitem{13}
  T. S. Monteiro, P. A. Dando, N. A. C. Hutchings, and M. R. Isherwood, Phys. Rev. Lett. \textbf{89}, 194102 (2002).

  \bibitem{14}
  I. Dana. V. Ramareddy, I. Talukdar, and G. S. Summy, Phys. Rev. Lett. \textbf{100}, 024103 (2008).

  \bibitem{15}
  M. Sadgrove, M. Horikoshi, T. Sekimura, and K. Nakagawa, Eur. Phys. J. D. \textbf{45}, 229 (2007).

  \bibitem{16}
  T. Salger, S. Kling, T. Hecking, C. Geckeler, L. Morales-Molina, and M. Weitz, Science \textbf{326}, 1241 (2009).

  \bibitem{17}
  R. D. Astumian and P. H\"{a}nggi, Phys. Today \textbf{55}, 33 (2002).

  \bibitem{18}
  M. Sadgrove, M. Horikoshi, T. Sekimura, and K. Nakagawa, Phys. Rev. Lett. \textbf{99}, 043002 (2007).

\bibitem{44}
  P. Reimann, M. Grifoni, and P. H\"{a}nggi, Phys. Rev. Lett. \textbf{79}, 10 (1997).



  \bibitem{21}
  E. Lundh and M. Wallin, Phys. Rev. Lett. \textbf{94}, 110603 (2005).

  \bibitem{22}
  A. Wickenbrock, D. Cubero, N. A. Abdul Wahab, P. Phoonthong, and F. Renzoni, Phys. Rev. E \textbf{84}, 021127 (2011).

  \bibitem{23}
  R. K. Shrestha, W. K. Lam, J. Ni, and G. S. Summy, Fluct. and Noise Lett. \textbf{12}, 1340003 (2013).

  \bibitem{24}
  R. K. Shrestha, J. Ni, W. K. Lam, S. Wimberger, and G. S. Summy, Phys. Rev. A \textbf{86}, 043617 (2012).

  \bibitem{25}
  Z.-Y. Ma, K. Burnett, M. B. d'Arcy, and S. A. Gardiner, Phys. Rev. A \textbf{73}, 013401 (2006).



  \bibitem{28}
  M. Kozuma, L. Deng, E. W. Hagley, J. Wen, R. Lutwak, K. Helmerson, S. L. Rolston, and W. D. Phillips, Phys. Rev. Lett. \textbf{82}, 871 (1999)


  \bibitem{29}
  H. Schanz, M.-F. Otto, R. Ketzmerick, and T. Dittrich, Phys. Rev. Lett. \textbf{87}, 070601 (2001).

\bibitem{42}
  S. Flach, O. Yevtushenko, and Y. Zolotaryuk, Phys. Rev. Lett. \textbf{84}, 2358 (2000).

 \bibitem{30}
  S. Wimberger, and M. Sadgrove, J. Phys. A: Math. Gen. \textbf{38}, 10549 (2005).

\bibitem{36}
  M. Sadgrove, and S. Wimberger, Adv. At. Mol. Opt. Phys. \textbf{60}, 315 (2011).

\bibitem{31}
  G. S. Summy, and S. Wimberger, Phys. Rev. A \textbf{93}, 023638 (2016).

  \bibitem{43}
  M. Weiss, C. Groiseau, W. Lam, R. Burioni, A. Vezzani, G. Summy, and S. Wimberger, Phys. Rev. A \textbf{92}, 033606 (2015).



\bibitem{34}
  T. Mazzoni, R. D. Aguila, L. Salvi, N. Poli, and G. M. Tino, Phys. Rev. A \textbf{92}, 053619 (2015).

\bibitem{37}
  M. D. Barrett, J. A. Sauer, and M. S. Chapman, Phys. Rev. Lett. \textbf{87}, 010404 (2001).

\bibitem{38}
  R. Graham, M. Schlautmann, and P. Zoller, Phys. Rev. A \textbf{45}, R19 (1992).


\bibitem{40}
  C. V. Raman and N. S. N. Nath, Proc. Ind. Acad. Sci. \textbf{4}, 222 (1936).
  
\bibitem{41}
  P. L. Gould, G. A. Ruff, and D. E. Pritchard, Phys. Rev. Lett. \textbf{56}, 827 (1986).
  

  
  
\end{thebibliography}
\end{document}